\documentclass[aps,prl,twocolumn,superscriptaddress,showpacs,floatfix,preprintnumbers]{revtex4-2}

\usepackage[dvipsnames]{xcolor}     
\definecolor{lcolor}{rgb}{0.5,0,0}
\definecolor{citcolor}{rgb}{0,0,1}
\usepackage[breaklinks,colorlinks,urlcolor=blue,citecolor=citcolor,linkcolor=lcolor,linktoc=all]{hyperref}
\usepackage{color}
\usepackage{graphicx}	
\graphicspath{{./figures/}}
\usepackage[utf8]{inputenc}
\usepackage{amsmath} 
\usepackage{amssymb}
\usepackage{slashed}
\usepackage{bm,bbm,bbold}
\usepackage{booktabs}
\usepackage{dcolumn}
\usepackage{dsfont}
\usepackage{MnSymbol} 
\usepackage{cancel}
\usepackage{booktabs}

\usepackage{feynmandiagrams} 

\makeatletter
\def\@fnsymbol#1{\ensuremath{\ifcase#1\or *\or \dagger\or \ddagger\or
   \mathsection\or \mathparagraph\or \|\or **\or \dagger\dagger
   \or \ddagger\ddagger \else\@ctrerr\fi}}
\makeatother

\allowdisplaybreaks

\definecolor{color0}{HTML}{648FFF} 
\definecolor{color1}{HTML}{785EF0} 
\definecolor{color2}{HTML}{DC267F} 
\definecolor{color3}{HTML}{FE6100} 
\definecolor{color4}{HTML}{FFB000} 

\makeatletter
\g@addto@macro\bfseries{\boldmath}
\makeatother

\usepackage{tikz, tikzscale}
\usetikzlibrary{calc, fadings, shadings, decorations.pathmorphing, decorations.pathreplacing, decorations.shapes}
\usepackage[customcolors]{hf-tikz}
\usepackage{mciteplus}

\begin{document}

\title{Perturbative QCD correction to the nonleptonic weak rate in dense quark matter}

\preprint{HIP-2026-16/TH}

\author{Matti Heikinheimo}
\email{matti.heikinheimo@helsinki.fi}
\affiliation{Department of Physics and Helsinki Institute of Physics,
P.O.~Box 64, FI-00014 University of Helsinki, Finland}
\author{Hanna Lempiäinen}
\email{hanna.lempiainen@helsinki.fi}
\affiliation{Department of Physics and Helsinki Institute of Physics,
P.O.~Box 64, FI-00014 University of Helsinki, Finland}
\author{Risto Paatelainen}
\email{risto.paatelainen@utu.fi}
\affiliation{Department of Physics and Astronomy, FI-20014 University of Turku, Finland}
\affiliation{Department of Physics and Helsinki Institute of Physics, P.O.~Box 64, FI-00014 University of Helsinki, Finland}
\email{risto.paatelainen@helsinki.fi}
\author{Tomi Ruosteoja}
\email{tomi.ruosteoja@helsinki.fi}
\affiliation{Department of Physics and Helsinki Institute of Physics,
P.O.~Box 64, FI-00014 University of Helsinki, Finland}
\author{Kimmo Tuominen}
\email{kimmo.i.tuominen@helsinki.fi}
\affiliation{Department of Physics and Helsinki Institute of Physics,
P.O.~Box 64, FI-00014 University of Helsinki, Finland}

\begin{abstract}
We present the first calculation of the leading perturbative quantum chromodynamics (QCD) correction
to the nonleptonic weak flavor-equilibration rate in dense quark matter. The computation is based on the Kadanoff--Baym kinetic equation and performed using real-time thermal field theory.
The resulting QCD correction is both gauge invariant and fully finite. In particular, all soft and collinear divergences cancel after combining real and virtual corrections, in accordance with the Kinoshita--Lee--Nauenberg theorem. In the low-temperature regime relevant to compact-star dynamics, the correction enhances the tree-level rate by $30 - 40\%$ over the high-density range where perturbative QCD is applicable.  The corrected rate improves on the tree-level result used in phenomenological calculations for more than three decades, providing a more accurate microscopic input for bulk-viscosity calculations in dense quark matter.
\end{abstract}

\maketitle

\emph{\textbf{Introduction}.}---%
Weak flavor-changing processes provide the microscopic mechanism for chemical equilibration in dense quark matter (QM)~\cite{Schmitt:2017efp}. These processes involve an interplay between weak interactions and quantum chromodynamics (QCD). Their rates are dynamical quantities that set the equilibration timescale and provide the microscopic input for calculating the bulk viscosity of quark matter~\cite{Alford:2017rxf, CruzRojas:2024etx,Hernandez:2025zxw}. The associated bulk-viscous dissipation can contribute to damping density oscillations in neutron stars and binary neutron-star mergers~\cite{Madsen:1993xx, Madsen:1998qb, Alford:2013pma, Alford:2010fd}.

In neutrino-transparent three-flavor QM, the dominant weak flavor-equilibration process is the nonleptonic reaction $u+d \rightleftarrows u+s$, mediated by the $W$ boson~\cite{Schmitt:2017efp}. Departures from chemical equilibrium are characterized by the chemical potential imbalance
$\delta\mu \equiv \mu_s - \mu_d$, which drives a net weak conversion between $d$ and $s$ quarks and induces changes in the corresponding quark densities, $\dot{n}_d=-\dot{n}_s\equiv\Gamma_d$. Here, the dot denotes a time derivative, and $\Gamma_d$ is the net rate of $s\to d$ conversions per unit volume and time. In the degenerate low-temperature limit $T\ll\mu$ with $\mu \equiv \mu_u = \mu_d$, and for small departures from equilibrium $|\delta\mu|\ll T$, the rate takes the linear-response form $\Gamma_d=\lambda_1\delta\mu$. 
The coefficient $\lambda_1$ enters directly into the calculation of the bulk viscosity of quark matter~\cite{CruzRojas:2024etx,Hernandez:2025zxw}.

Despite its importance, the rate $\Gamma_d$ has so far been known only at tree level, based on thermal field theory calculations dating back more than three decades~\cite{Madsen:1993xx, Heiselberg:1992bd}. In the limit $|\delta\mu| \ll T \ll \mu$, neglecting small quark-mass corrections, the rate at leading power in $T/\mu$ is~\cite{Madsen:1993xx}
\begin{equation}
\Gamma_d^{(0)} 
=
\frac{64}{45\pi^3}\,N_c^2\,
G_F^2\,
|V_{ud}|^2|V_{us}|^2\,
\mu^5\,T^2\,\delta\mu.
\label{eq:GammaLOIntro}
\end{equation}
Here, the superscript $(0)$ denotes the weak tree-level contribution, $G_F\equiv g_W^2/(4\sqrt{2}M_W^2)$ is the Fermi coupling, $N_c = 3$ the number of colors, and $V_{ud}$ and $V_{us}$ are the relevant Cabibbo--Kobayashi--Maskawa (CKM) matrix elements~\cite{Kobayashi:1973fv}. Higher-order corrections, in particular those governed by the strong coupling $\alpha_s\equiv g^2/(4\pi)$, have remained unknown. By contrast, systematic next-to-leading-order calculations of real-time rates and transport coefficients in hot QCD (see e.g.~\cite{Caron-Huot:2007rwy,Ghiglieri:2013gia,Ghiglieri:2018dib,Ghiglieri:2015ala,Wu:2024vyc}) have shown that perturbative corrections can significantly modify leading-order predictions.

In this Letter and the accompanying paper~\cite{companionpaper}, we use the Kadanoff--Baym (KB) formalism of real-time thermal field theory~\cite{Kadanoff:1989xx,Ghiglieri:2020dpq} to calculate the leading perturbative QCD correction at $O(G_F^2\alpha_s)$ to the nonleptonic weak flavor-equilibration rate $\Gamma_d$ in dense, unpaired quark matter. The explicit form of the QCD correction depends on the hierarchy between the temperature $T$ and the gluon screening mass $m_{\rm E}\sim\sqrt{\alpha_s}\,\mu$ in dense quark matter~\cite{Gorda:2021kme}.

In the phenomenologically interesting low-temperature regime $T\ll m_{\rm E}\ll\mu$, the QCD correction is sensitive to the screening scale and contains a logarithm of $m_{\rm E}/\mu\sim\sqrt{\alpha_s}$. Its evaluation requires Hard--Dense--Loop (HDL) resummation of soft gluonic modes~\cite{Braaten:1989mz, Braaten:1991gm,Manuel:1995td,Ipp:2003cj}. In the opposite hierarchy, $m_{\rm E}\ll T\ll\mu$, the thermal scale sets the relevant infrared scale, yielding instead a logarithm of $T/\mu$. At the order considered, the result can be obtained using bare (unresummed) gluon exchange. In both regimes, the QCD correction is gauge invariant and free of ultraviolet and infrared singularities. Soft and collinear divergences cancel between real and virtual contributions in accordance with the Kinoshita--Lee--Nauenberg (KLN) theorem~\cite{Kinoshita:1962ur, Lee:1964is}.

\emph{\textbf{Theoretical Setup}.}—%
Near equilibrium, the net down-quark production rate is expressed via the KB kinetic equation as~\cite{Schmitt:2025cqi, companionpaper},
\begin{equation}
\Gamma_d = i\int_P \mathrm{Tr}
\Bigl[
\Sigma_d^{<}(P)\, S_d^{>}(P)
-
\Sigma_d^{>}(P)\,S_d^{<}(P)
\Bigr] \, ,
\label{eq:GammauKB}
\end{equation}
where $S^{\gtrless}_d$ and $\Sigma^{\gtrless}_d$ are the backward ($<$) and forward ($>$) Wightman propagators and self-energies of the down quark, respectively. The self-energies encode the interactions of the propagating down quark with the medium, while 
Eq.~\eqref{eq:GammauKB} has the familiar gain-loss structure of kinetic theory: the first term describes the gain of down quarks and the second their loss through the nonleptonic process $u + d \;\rightleftarrows\; u + s$; their difference yields the net down-quark production rate. We use mostly plus metric, the notation $P=(p^0,\mathbf{p})$ for the four-momentum of the propagating down quark, $\int_P \equiv \int \mathrm{d}^4P/(2\pi)^4$, and the trace is taken over color and Dirac indices.  

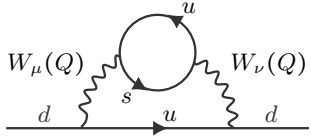
\begin{figure}[h!]
    \centering
        \begin{tikzpicture}
            \draw[->-,->-/node=A] (-2,0) 
            -- (-1.5,0) node[anchor=south, inner sep=2pt] {$d$} 
            -- (1.5,0) node[anchor=south,inner sep=2pt] {$d$} 
            -- (2,0);
            \draw[snake] (1,0) arc (0:180:1);
            \draw[->->-,->->-/pos1=0.167,->->-/pos2=0.667,->->-/node1=B,->->-/node2=C,fill=white] (0,1) circle (0.5);
            \node[anchor=-135,inner sep=2.5pt] at (A) {$u$};
            \node[anchor=-135,inner sep=2.5pt] at (B) {$u$};
            \node[anchor=45,inner sep=2.5pt] at (C) {$s$};
            \node at (150:1.7) {$W_{\mu}(Q)$};
            \node at (30:1.7) {$W_{\nu}(Q)$};
        \end{tikzpicture}
    \caption{The $d$-quark self-energy, which contributes to the net $d$-quark production at tree-level in the weak coupling.     
    }
    \label{fig:SigmadLO}
\end{figure}

In Eq.~\eqref{eq:GammauKB}, flavor-conserving contributions appear identically in the gain and loss terms and therefore cancel, leaving only flavor-changing self-energies. At leading order in the weak interaction, the surviving contribution corresponds to the self-energy shown in Fig.~\ref{fig:SigmadLO}, where a down quark exchanges a $W$ boson whose polarization tensor contains
a one-loop $u$--$s$ quark bubble. 

Inserting this self-energy into Eq.~\eqref{eq:GammauKB} and using
$Q^2\ll M_W^2$, one obtains 
\begin{equation}
\begin{split}    
\Gamma_d^{(0)} = -\frac{1}{M_W^4} \!  \int_Q & \Bigl \{
\textrm{Tr}[\Pi^{(0) \, >}_{d}(Q) \,
\Pi^{(0) \, <}_{s}(Q)] \\
& - \textrm{Tr}[\Pi^{(0) \, <}_{d}(Q) \, \Pi^{(0) \, >}_{s}(Q)] \Bigr \}  \,,
\end{split}
\label{eq:GammadLO}
\end{equation}
where the trace denotes the contraction of Lorentz indices, $\textrm{Tr}[\Pi_{d} \, 
\Pi_{s}] \equiv \Pi_{d,\mu\nu} 
\Pi^{\nu\mu}_{s}$. The corresponding one-loop charged-current polarization tensor for $f=d,s$ is
\begin{equation}
-i\Pi^{(0) \, \gtrless}_{f,\mu\nu} = \frac{g_W^2 N_c|V_{uf}|^2}
{2} \!\!\!\int_R
\mathrm{Tr}
\Bigl[
S_f^{\gtrless}(R) \Gamma_\mu 
S_u^{\lessgtr}(R-Q) \Gamma_\nu
\Bigr] \, ,
\label{eq:WpolLO}
\end{equation}
where $\Gamma_\mu \equiv \gamma_\mu(1-\gamma^5)/2$. Because the charged weak current contains both vector and axial-vector components, the corresponding polarization tensor contains both symmetric and antisymmetric Lorentz structures.

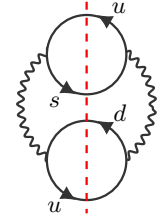
\begin{figure}[h!]
    \centering
    \begin{tikzpicture}[scale=0.7]
        \draw[snake] (0,0) circle (1.25);
        \draw[->->-,->->-/pos1=0.167,->->-/pos2=0.667,->->-/node1=A,->->-/node2=B,fill=white] (0,1) circle (0.75);
        \draw[->->-,->->-/pos1=0.167,->->-/pos2=0.667,->->-/node1=C,->->-/node2=D,fill=white] (0,-1) circle (0.75);
        \node[anchor=-135,inner sep=2.5pt] at (A) {$u$};
        \node[anchor=45,inner sep=2.5pt] at (B) {$s$};
        \node[anchor=-135,inner sep=2.5pt] at (C) {$d$};
        \node[anchor=45,inner sep=2.5pt] at (D) {$u$};
        \draw[cut] (0,2) -- (0,-2);
    \end{tikzpicture}
    \caption{Diagrammatic representation of the tree-level contribution to $\Gamma_d$, expressed as a cut (red dashed line) through the charged-current polarization tensors.}
    \label{fig:GammadLOspectral}
\end{figure}

Near equilibrium, the charged-current Wightman self-energies satisfy the generalized Kubo-Martin-Schwinger (KMS) relation $\Pi_f^{>}(Q)
=
e^{\beta(q^0-\delta\mu_f)}
\Pi_f^{<}(Q)$, where $\beta\equiv1/T$ and assuming the general case where the chemical-potential imbalance $\delta\mu_f$ associated with the flavor-changing current is $\delta\mu_f\equiv\mu_f-\mu_u \neq 0$. Consequently, Eq.~\eqref{eq:GammadLO} can be rewritten in terms of
$\Pi_f^{-}\equiv\Pi_f^{<}-\Pi_f^{>}=\Pi_f^{R}-\Pi_f^{A}$,
where the superscripts $R$ and $A$ denote the retarded and advanced charged-current self-energies, respectively. This yields the tree-level expression
\begin{equation}
\begin{split}    
\Gamma_d^{(0)} = -\frac{1}{M_W^4} \!  \int_Q & \Bigl \{n_B(q^0 - \delta\mu_s) - n_B(q^0 - \delta\mu_d)\Bigr \} \\
& \times \textrm{Tr}[\Pi^{(0) \, -}_{d}(Q) \, \Pi^{(0) \, -}_{s}(Q)]  \,, \\
&
\end{split}
\label{eq:GammadLOspectral}
\end{equation}
where $n_B(x)\equiv(e^{\beta x}-1)^{-1}$ is the Bose-Einstein distribution. Figure~\ref{fig:GammadLOspectral} illustrates the diagrammatic representation of Eq.~\eqref{eq:GammadLOspectral}, in which each factor $\Pi_f^{-}$ is represented by a cut charged-current self-energy, placing the internal quark lines on shell.

In the degenerate regime where $\vert \delta\mu_f \vert \ll \mu$, the Bose factors restrict the energy transfer to $\vert q^0 \vert \ll\mu$. Decomposing the charged-current spectral functions into transverse, longitudinal, and antisymmetric components (see~\cite{companionpaper}), the remaining integrations can be performed analytically. 
In the limit $|\delta\mu|=|\delta\mu_s-\delta \mu_d|\ll T$ the result is 
the tree-level rate given in Eq.~\eqref{eq:GammaLOIntro}.

The main advantage of the KB formulation in
Eq.~\eqref{eq:GammauKB} is that it provides a systematic framework for organizing perturbative corrections to $\Gamma_d$. The down-quark propagator and self-energy admit perturbative expansions in both the strong and weak interactions, and their relevant parts, contributing to the rate  $\Gamma_d$ at $O(G_F^2\alpha_s)$, are shown in Eqs.~\eqref{eq:gquarkpro} and \eqref{eq:gselfkpro}.  They consist of pure-QCD contributions, weak contributions, and combined weak--QCD contributions. The gluon propagators appearing in these diagrams are fully HDL-resummed, accounting for the modification of gluon propagation in the dense quark medium, with the relevant soft scale set by $m_{\rm E}$. Note that diagrams with a single gluon attached to a closed quark loop containing two $W$-boson vertices vanish identically because the associated color trace vanishes.

\begin{widetext}
\begin{equation}
S_d = \quad \diagQuarkPropBare[0.6] +   \diagQuarkPropWW[0.6] + \diagQuarkPropGG[0.6] + \cdots
\label{eq:gquarkpro}
\end{equation}
\begin{equation}
\begin{split}
    -i\Sigma_d =  \quad  &  \diagQuarkSeWW[0.6] + \diagQuarkSeGG[0.6] + \diagQuarkSeGWW[0.6] + \diagQuarkSeWGW[0.6]+ \diagQuarkSeWWG[0.6] +   \\
    & \diagQuarkSeWGGW[0.6] +  \diagQuarkSeWGWG[0.6] 
     + \diagQuarkSeGWGW[0.6] +        
\diagQuarkSeGWWG[0.6] +\diagQuarkSePiBlob[0.6] + \cdots
\end{split}
\label{eq:gselfkpro}
\end{equation}
\end{widetext}

The corresponding QCD correction to the charged-current polarization tensor in Eq.~\eqref{eq:WpolLO} is given by the two-loop W-boson self-energy,
$\Pi^{(1)}=\Pi_{\mathrm{Ia}}+\Pi_{\mathrm{Ib}}+\Pi_{\mathrm{II}}$:
\begin{equation}
\begin{split}
      -i\Pi_{\text{Ia}} & = \diagWSeIa[0.6],\quad 
      -i\Pi_{\text{Ib}}  = \diagWSeIb[0.6], \\ 
      -i\Pi_{\text{II}} & = \diagWSeII[0.6] \, .
    \end{split}
\end{equation}
Here, the contributions Ia and Ib correspond to quark self-energy
insertions, while contribution II is the vertex correction.

At $O(G_F^2\alpha_s)$, the compact Wightman self-energy representation of the
tree-level rate in Eq.~\eqref{eq:GammadLO} is not manifest diagram by
diagram. The perturbative expansion of the KB collision term generates a
large number of distinct real-time contributions. After summing the gain and loss terms, nontrivial cancellations occur among these contributions (see~\cite{companionpaper} for details),  leaving the remarkably compact expression
\begin{equation}
\label{eq:QCD rate with W self-energy}
\begin{split}
    \Gamma_d^{(1)} = -\frac{1}{M_W^4}&\int_{Q} \Bigl \{
        \textrm{Tr}[\Pi_{d}^{(0) \, >} \, \Pi_{s}^{(1)\,<} + \Pi_{d}^{(1) \, >} \, \Pi_{s}^{(0) \, <}] \\
        &- \textrm{Tr}[\Pi_{d}^{(0) \, <} \, \Pi_{s}^{(1) \, >} + \Pi_{d}^{(1) \, <} \, \Pi_{s}^{(0) \, >}]\Bigr \} \,.
\end{split}
\end{equation}

Applying the generalized KMS relations, this Wightman-function representation can be rewritten as 
\begin{equation}
\label{eq:QCD rate with W self-energy}
\begin{split}
    \Gamma_d^{(1)} = -\frac{1}{M_W^4}&\int_{Q}  \Bigl \{n_B(q^0 - \delta\mu_s) - n_B(q^0 - \delta\mu_d)\Bigr \} \\
    & \times 
        \textrm{Tr}[\Pi_{d}^{(0) \, -}\, \Pi_{s}^{(1)\,-}+ \Pi_{d}^{(1) \, -} \, \Pi_{s}^{(0) \, -}] \,.
\end{split}
\end{equation}

To evaluate Eq.~\eqref{eq:QCD rate with W self-energy}, we isolate the gluon momentum $K$ and write the correction as
\begin{equation}
    \Gamma_d^{(1)} =
    \int_K D_{\alpha\beta}\,F^{\alpha\beta},
    \label{eq:Gamma_NLO_schematic}
\end{equation}
where the explicit real-time indices in the $R/A$ basis are suppressed. Here, $D_{\alpha\beta}= D_{\alpha\beta}(K)$ denotes the resummed HDL gluon propagator, while $F^{\alpha\beta}$ contains the remaining part of the integrand.

Defining
\begin{equation}
    D_I(K) =
    \frac{-i}{K^2+m_{\rm E}^2\Pi^{\rm HDL}_I(k^0/k)}, 
\label{eq:DI_definition}
\end{equation}
where $\Pi^{\rm HDL}_I$ denotes the transverse ($I = {\rm T}$) and longitudinal ($I= {\rm L}$) components of the soft HDL gluon  self-energies \cite{Gorda:2023zwy}, the resummed gluon propagator in covariant $R_\xi$ gauge can
be decomposed as~\cite{companionpaper}
\begin{equation}
\begin{split}
    D_{\alpha\beta}(K)
    & =
    D_{\rm T}(K)g_{\alpha\beta} +
    \frac{K^2}{k^2}
    \bigl[D_{\rm T}(K)-D_{\rm L}(K)\bigr]
    N_\alpha N_\beta \\
    & + a(K) (K_\alpha N_\beta + K_\beta N_\alpha) + b(K,\xi) K_\alpha K_\beta.    \label{eq:gluon_decomposition}
\end{split}
\end{equation}
Here, $N_{\alpha}=(1,\mathbf{0})$, and $a, b$ are scalar coefficients, with $b$ depending explicitly on the gauge-fixing parameter $\xi$.
The last two terms do not contribute to $\Pi_f^-$ and, consequently, do not affect the rate.
Hence,
\begin{equation}  
    \Gamma_d^{(1)}  =
    \int_K
    \biggl [
         D_{\rm T}\,F_\alpha^{\,\,\alpha}  +               \frac{K^2}{k^2}
        \bigl[D_{\rm T}-D_{\rm L}\bigr]\,
        F^{00}
    \biggr ].
    \label{eq:Gamma_NLO_reduced}
\end{equation}

To evaluate the QCD correction to the rate in Eq.~\eqref{eq:Gamma_NLO_reduced}, we first consider the low-temperature regime $T\ll m_{\rm E}$, which is most relevant for phenomenological applications.
Direct evaluation of the above expression requires retaining both the full HDL dependence and the nontrivial momentum dependence of $F^{\alpha\beta}$. We circumvent this by adding and subtracting the light-cone limit of the transverse HDL propagator, $D_\infty(K)\equiv-i/(K^2+m_\infty^2)$, where $m_\infty$ is the asymptotic thermal mass of the transverse gluon mode, given by $m_\infty^2 = m_{\rm E}^2/2$. We obtain
\begin{align}
    \Gamma_d^{(1)}
    =
    \int_K
    \Biggl[
        D_\infty\,
        F_\alpha^{\,\,\alpha}
        & +
        \bigl[D_{\rm T}-D_\infty\bigr]\,
        F_\alpha^{\,\,\alpha}
        \nonumber\\
        &+
        \frac{K^2}{k^2}
        \bigl[D_{\rm T}-D_{\rm L}\bigr]\,
        F^{00}
    \Biggr].
    \label{eq:Gamma_NLO_subtracted}
\end{align}
The combinations $D_{\rm T}-D_\infty$ and $(K^2/k^2)[D_{\rm T}-D_{\rm L}]$ are suppressed for hard $K\sim\mu$ and vanish on the light cone. Thus, to the required order in $m_{\rm E}/\mu$, the hard region does not contribute to the terms retaining the full HDL dependence, allowing $F^{\alpha\beta}$ to be expanded in $K/\mu$. The remaining term retains the full momentum dependence of $F^{\alpha\beta}$ but involves only the simpler propagator $D_\infty$.

We compute this expression in dimensional regularization with $D=4-2\epsilon$, treating $\gamma_5$ in Larin's scheme~\cite{Larin:1993tq}. The first term in Eq.~\eqref{eq:Gamma_NLO_subtracted} is computed analytically in the limit $m_{\rm E}\ll\mu$, while in the remaining terms $F^{\alpha\beta}$ is expanded in the soft gluon momentum. After performing all other integrations analytically, the remaining integral over $x\equiv k^0/k$, containing the full HDL dependence, is evaluated numerically.

In the opposite regime $m_{\rm E}\ll T\ll\mu$, the temperature sets the relevant infrared scale, and HDL resummation is not required at the order considered. We can therefore take the $m_{\rm E}\to 0$ limit of Eq.~\eqref{eq:Gamma_NLO_reduced}, which ultimately corresponds to replacing the gluon propagator in Eq.~\eqref{eq:Gamma_NLO_schematic} with the bare propagator $D_{\alpha\beta}(K) = -ig_{\alpha\beta}/K^2$. The resulting expression can be evaluated fully analytically.

In both regimes, ultraviolet divergences cancel between the self-energy and vertex contributions. The soft and collinear singularities cancel between the gluon-emission and virtual-exchange cuts, separately within the type-I and type-II contributions. For the inclusive rate $\Gamma_d$, these real–virtual cancellations are the finite-temperature and finite-density realization of the KLN theorem.

\emph{\textbf{Results}.}---%
We now present the QCD correction to the tree-level flavor-equilibration rate in Eq.~\eqref{eq:GammaLOIntro}, starting with the low-temperature \(T\ll m_{\rm E}\ll\mu\) regime relevant for phenomenological applications. In this regime, the result reads
\begin{equation}    
\Gamma_{d}^{(1)}=\Gamma_d^{(0)}
\frac{\alpha_s C_F}{\pi}
\left(A\ln\frac{m_{\rm E}}{\mu}+B\right)
+O(\alpha_s^2),
\label{eq:mainresult1}
\end{equation}
where the color factor $C_F=(N_c^2-1)/(2N_c)$ and $m_{\rm E}^2/\mu^2=(2/\pi)N_f\alpha_s$ with $N_f$ being the number of quark flavors. The coefficients $A$ and $B$ are given by:
\begin{equation}
\begin{split}    
A= & -\frac{541}{60}+8\ln2,\\
B  =  & -\frac{36137}{3600}
 -\frac{15\pi^2}{128}
+\frac{619}{24}\ln2 
-12\ln^22 + 1.37064. 
\end{split}    
\end{equation}
The numerical part in the coefficient $B$ contains an angular integral over HDL self-energies that admits no closed-form expression and is therefore evaluated numerically to high precision. For $N_c = N_f=3$, the result above can be written numerically as
$\Gamma^{(1)}_d/\Gamma_d^{(0)}
\simeq
\alpha_s\left(-0.7366\,\ln\alpha_s + 0.4943\right)$,
making the non-analytic $\alpha_s\ln\alpha_s$ dependence explicit.

For the opposite hierarchy, \(m_{\rm E}\ll T\ll\mu\), the temperature regulates the infrared region and HDL resummation is not required at the order considered. The result can therefore be obtained using bare gluon exchange and reads
\begin{equation}    
\Gamma_{d}^{(1)}=\Gamma_d^{(0)}\frac{\alpha_s C_F}{\pi}
\left(A\ln\frac{e^{-\gamma_E}\pi T}{\mu}+\widetilde B\right)
+O(\alpha_s^2),
\label{eq:mainresult2}
\end{equation}
where $\gamma_E \simeq 0.577$ is the Euler-Mascheroni constant and the coefficient $\widetilde B$ is obtained analytically,
\begin{align}
\widetilde B=-\frac{6479}{1200}
-\frac{15\pi^2}{128}
-\frac{2}{3}\ln2
+4\ln^22.
\end{align}

These two results reflect the change in the infrared dynamics as the temperature is lowered. In the regime $m_{\rm E}\ll T\ll\mu$, the logarithmic correction in Eq.~\eqref{eq:mainresult2} grows as $T/\mu$ decreases, signaling increasing sensitivity to soft gluon exchange. For $T\ll m_{\rm E}$, the soft momentum region is dominated by screened gluon exchange, and the naive perturbative treatment breaks down. The soft gluon propagator must therefore be HDL-resummed, with the screening scale $m_{\rm E}$ replacing the temperature as the scale entering the logarithmic correction as in Eq.~\eqref{eq:mainresult1}.

\begin{figure}[t]
    \centering
    \includegraphics[width=\columnwidth]{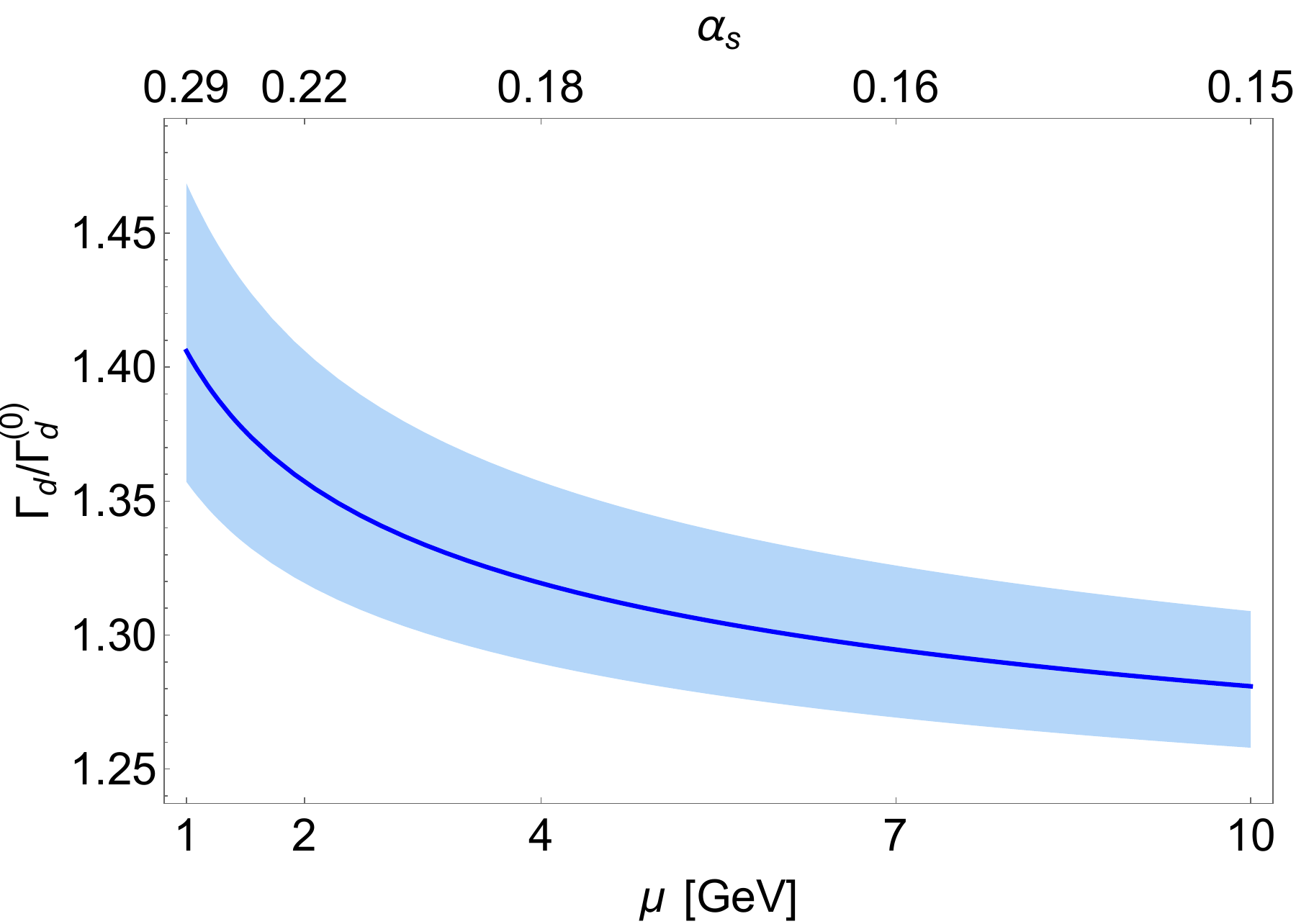}
\caption{Ratio of the QCD-corrected weak flavor-equilibration rate to the tree-level result as a function of the quark chemical potential $\mu$. The solid line shows the central result obtained with the renormalization scale $\bar{\Lambda}=2\mu$, while the shaded band is obtained by varying the scale over $\bar{\Lambda}\in[\mu,4\mu]$. The upper axis shows the corresponding values of $\alpha_s$ at the central renormalization scale.}
    \label{fig:plot}
\end{figure}

Figure~\ref{fig:plot} shows the total rate  normalized to the tree-level result, $\Gamma_{d}/\Gamma_{d}^{(0)}=1+\Gamma_{d}^{(1)}/\Gamma_{d}^{(0)}$, where $\Gamma_{d}^{(1)}$ is evaluated using the low-temperature result in Eq.~\eqref{eq:mainresult1}. We use the one-loop $\overline{\text{MS}}$ running coupling $\alpha_s = \alpha_s(\bar\Lambda) $~\cite{Kurkela:2009gj}, with central scale $\bar\Lambda=2\mu$, and vary the scale over $\bar\Lambda\in[\mu,4\mu]$. The QCD correction enhances the rate by approximately $30 - 40 \%$ over the range shown in Fig.~\ref{fig:plot}. The renormalization-scale variation leaves this enhancement intact, while its magnitude decreases gradually toward weaker coupling.

Finally, we comment on previously proposed non-Fermi-liquid enhancements of the nonleptonic weak reaction rate in dense QM~\cite{Schwenzer:2012ga},
obtained by incorporating interaction-modified quasiparticle dispersion relations into the tree-level Boltzmann collision term with on-shell
external quarks. The resulting enhancement has subsequently been used in phenomenological calculations of the bulk viscosity~\cite{CruzRojas:2024etx}. The main result in this Letter
Eq.~\eqref{eq:mainresult1} exhibits no such enhancement at $O(G_F^2\alpha_s)$. As shown in the companion paper, this
discrepancy arises because the tree-level on-shell Boltzmann collision
term cannot be consistently modified by simply replacing the free dispersion relation with the quasiparticle one.

\emph{\textbf{Conclusions}.}—%
We have computed the leading perturbative QCD correction to the weak flavor-equilibration rate arising from the nonleptonic process $u + d \rightleftarrows u + s$ in dense quark matter, superseding the tree-level result used in phenomenological calculations for more than three decades. In the low-temperature regime relevant to compact-star applications, the correction enhances the tree-level rate by approximately $30 - 40\%$. The correction remains positive under scale variation and decreases gradually toward weaker coupling.

More broadly, our calculation establishes a systematic first-principles framework for computing weak equilibration rates in dense quark matter beyond tree level. The same framework can be applied to compute perturbative QCD corrections to semileptonic weak rates~\cite{Schmitt:2017efp}.
In color-superconducting phases, the quark propagators must instead be formulated in the Nambu--Gorkov basis to incorporate pairing-induced gaps~\cite{Schmitt:2025cqi,Wang:2009if}.

Finally, our result provides a perturbative high-density benchmark complementary to microscopic nuclear-matter calculations and strong-coupling holographic descriptions of weak flavor equilibration (see  e.g.~\cite{Schmitt:2017efp,Hoyos:2024pkl}). Together, these approaches could help map weak flavor-equilibration rates across the density range relevant to compact stars.

\emph{\textbf{Acknowledgments}.}---We thank Niko Jokela and Aleksi Vuorinen for useful discussions. This work has been funded by the Research Council of Finland (RCF) through projects 354533, 347499, 371542, and 353772, as well as by the Centre of Excellence in Neutron-Star Physics (project 374062). HL thanks the Finnish Cultural Foundation and TR the Vilho, Yrjö and Kalle Väisälä Foundation for financial support. RP has also been funded by the European Union (ERC, ExPertQCD, grant No. 101231521). Views and opinions expressed are, however, those of the author(s) only and do not necessarily reflect those of the European Union or the European Research Council. Neither the European Union nor the granting authority can be held responsible for them.

\bibliography{biblio.bib}

\end{document}